\documentclass[aps,prd,amsmath,amssymb,nofootinbib,tightenlines,floatfix,superscriptaddress]{revtex4}
\usepackage{epsfig}
\usepackage{latexsym}
\usepackage{amsmath}
\usepackage{amssymb}
\usepackage{amsthm}
\usepackage{amscd}
\usepackage{amsfonts}
\usepackage{graphicx}
\usepackage{fancyhdr}
\usepackage{url}
\usepackage{hyperref}
\usepackage{ifpdf}
\usepackage{epstopdf}
\begin{document}

\title{Prospects For Precision Measurements with Reactor Antineutrinos at Daya Bay}
\author{Daya Bay Collaboration}
\date{September 27, 2013}                                           

\begin{abstract}
In 2012 the Daya Bay experiment made an unambiguous observation of reactor antineutrino disappearance over kilometer-long baselines and determined that the neutrino mixing angle $\theta_{13}$ is non-zero. The measurements of Daya Bay have provided the most precise determination of $\theta_{13}$ to date. This whitepaper outlines the prospects for precision studies of reactor antineutrinos at Daya Bay in the coming years. This includes precision measurements of sin$^2$2$\theta_{13}$ and $\Delta m^2_{ee}$ to $<$3\%, high-statistics measurement of reactor flux and spectrum, and non-standard physics searches.  
\end{abstract}
\maketitle

\section{Introduction}
Since the discovery of a non-zero value for $\theta_{13}$~\cite{DYB1}, Daya Bay has provided the most precise measurement of sin$^2$2$\theta_{13}$ of $ 0.089\pm0.010\pm0.005$ with 139 days of 6 antineutrino detector (AD) data~\cite{DYB2} in June 2012. Recently, with 217 days of 6-AD data, we have obtained the most precise measurement of sin$^2$2$\theta_{13} = 0.090^{+0.008}_{-0.009}$ and the first measurement of
$\Delta m^2_{ee} = (2.59^{+0.19}_{-0.20}) \times 10^{-3}~eV^2$.

Two more ADs were installed and have been in operation since October 19, 2012. We present the expected precision in sin$^2$2$\theta_{13}$ and $\Delta m^2_{ee}$ as a function of time 
in Sections~\ref{sec:13} and \ref{sec:ee} respectively. The prospects for precise measurement 
of the reactor antineutrino flux and spectrum are summarized in Section \ref{sec:reactor}.

\section{Precision Measurement of sin$^2$2$\theta_{13}$}\label{sec:13}
\begin{figure}[htbp]
\begin{center}
\includegraphics[scale=0.7]{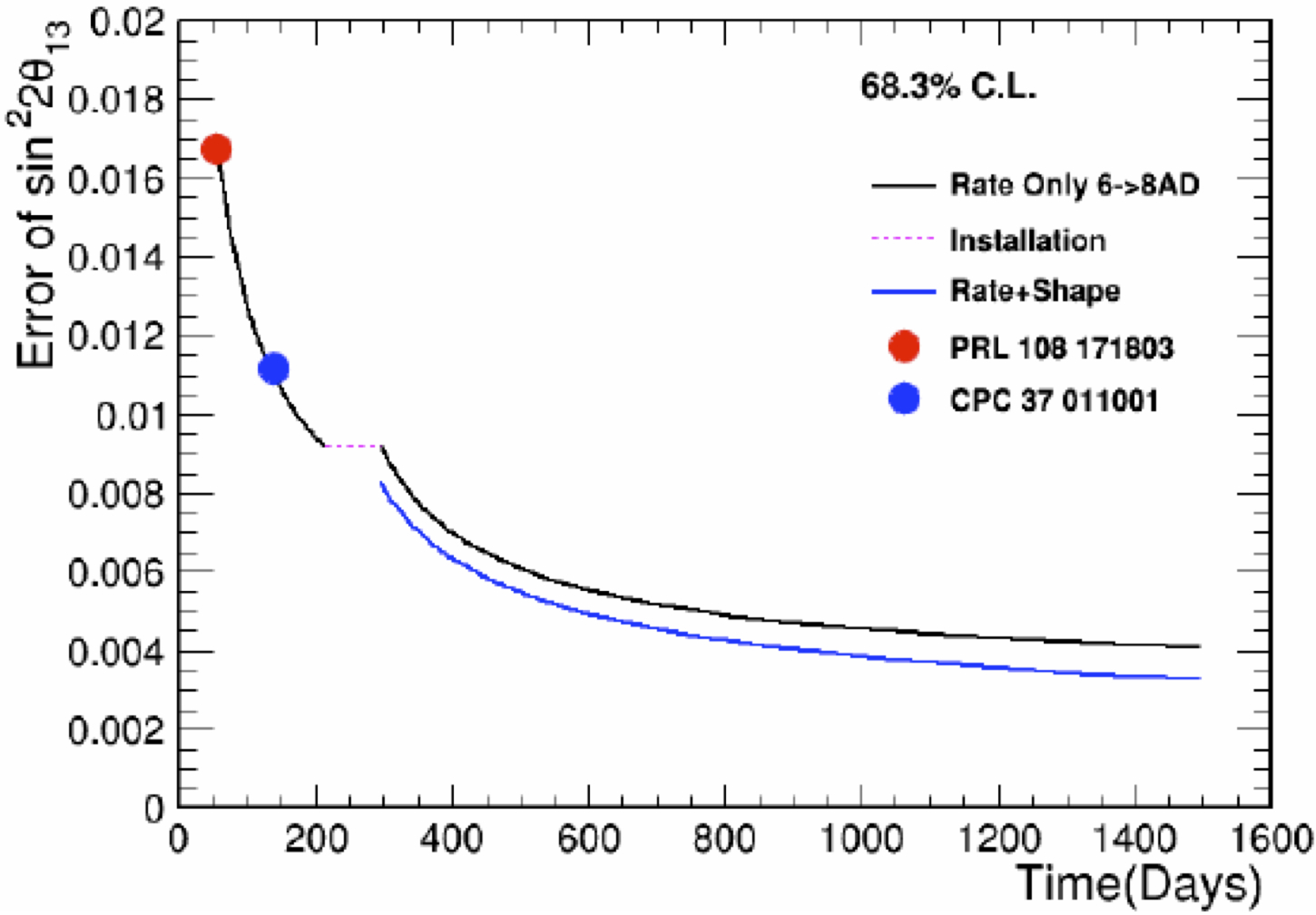}
\caption{Projected total uncertainty in sin$^2$2$\theta_{13}$ for the Daya Bay experiment.
The current systematic uncertainty of 0.005 is used at the beginning of the 8AD run.}
\label{fig:13}
\end{center}
\end{figure}

Daya Bay's currently published measurements of sin$^2$2$\theta_{13}$ are still statistics limited. With the data collected with 8 ADs as of October 2013 the statistical error will reach the current systematic uncertainty of 0.005~\cite{DYB2}. 
We have assessed improvements in the systematic uncertainty based on calibration data and increased precision that is statistical in nature (such as the uncertainties in estimating the $^9$Li/$^8$He and fast neutron backgrounds). 
Based on the projected systematic uncertainty we estimate that the total uncertainty in 
sin$^2$2$\theta_{13}$ can be reduced from 0.008 to 0.006 in one year and 0.003 after 3 
to 4 years as shown in Fig.~\ref{fig:13}.

\begin{figure}[htbp]
\begin{center}
\includegraphics[scale=0.5]{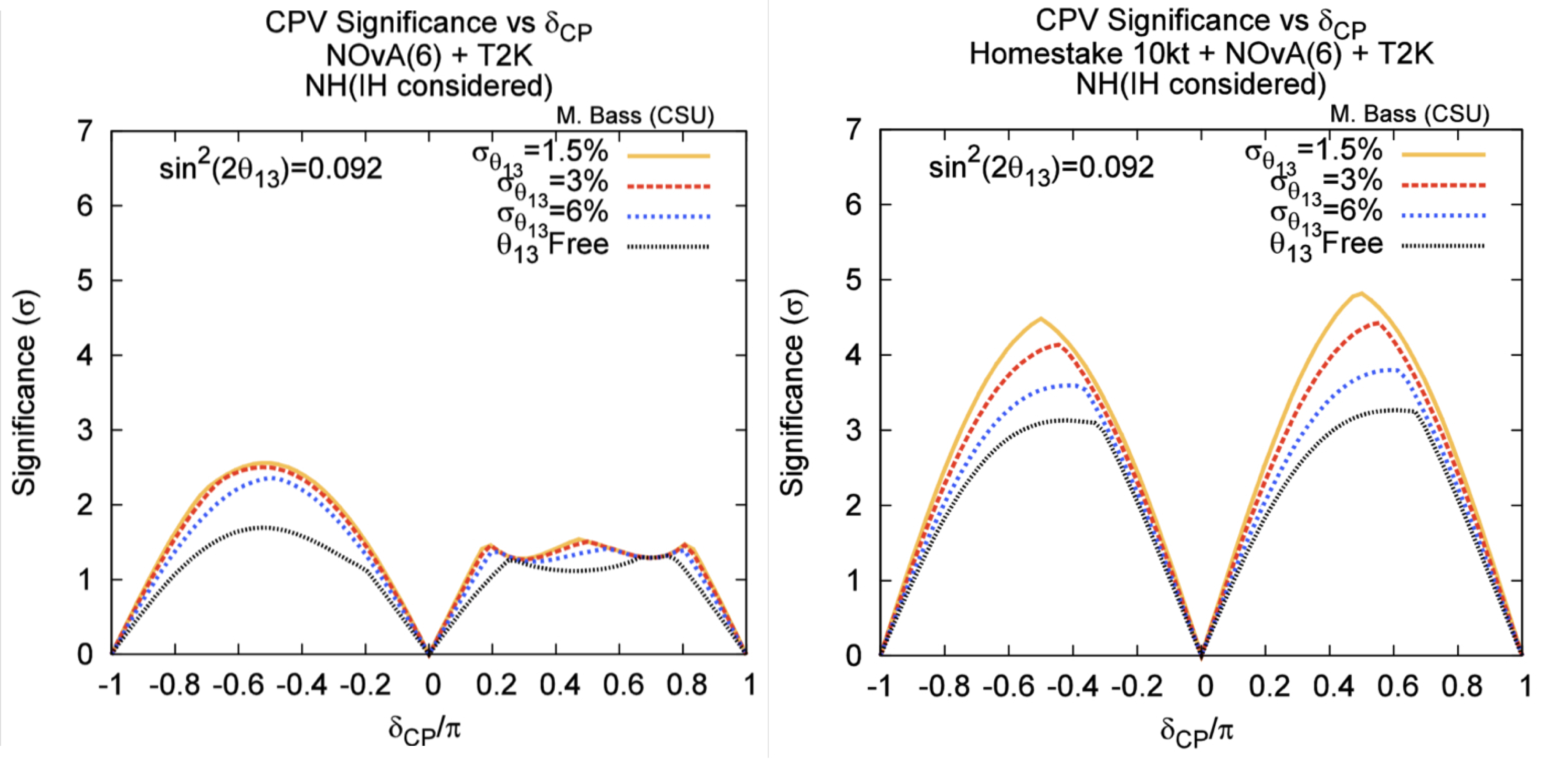}
\caption{Significance with which CP violation can be observed, by NOvA+T2K (left) and NOvA+T2K+LBNE (right), as a function of the value of $\delta_{CP}$. Observation of CP violation is equivalent to measuring $\delta_{CP}\neq 0, \pi$. The significance is calculated by minimizing over normal and inverted hierarchies, as the hierarchy is assumed to be unknown. The impact 
of $\theta_{13}$ precision is shown.}
\label{fig:CP}
\end{center}
\end{figure}

Daya Bay's short-baseline antineutrino disappearance measurement is the most precise reactor measurement of sin$^2$2$\theta_{13}$ and will likely remain the most precise measurement of this fundamental parameter for the foreseeable future. Precise measurement of this quantity may shed light on symmetries between quarks and leptons at a fundamental level, and test unitarity of the neutrino mixing matrix. Comparison of future long baseline accelerator measurements of $\theta_{13}$ to Daya Bay will allow precision tests of the neutrino-standard model interpretation of neutrino oscillations with sensitivity to non-standard neutrino interactions and sterile neutrino scenarios. The expected improvement in sin$^2$2$\theta_{13}$ by Daya Bay will make $\theta_{13}$ the most precisely measured neutrino mixing angle and extend the CP reach 
(shown in Fig.~\ref{fig:CP}) and $\theta_{23}$ octant determination 
of long baseline accelerator experiments. 

\section{Precision Measurement of $\Delta m^2_{ee}$}\label{sec:ee}
Our discovery of a large value for $\theta_{13}$ enables a measurement of the effective 
mass splitting $\Delta m^2_{ee}$ with a precision comparable to the MINOS measurement of 
$\Delta m^2_{\mu\mu}$. The disappearance of electron antineutrinos over km-long baselines observed at Daya Bay is a combination of antineutrino oscillations with mass splittings 
$\Delta m^2_{31}$ and $\Delta m^2_{32}$. 

\begin{figure}[htbp]
\begin{center}
\includegraphics[scale=0.8]{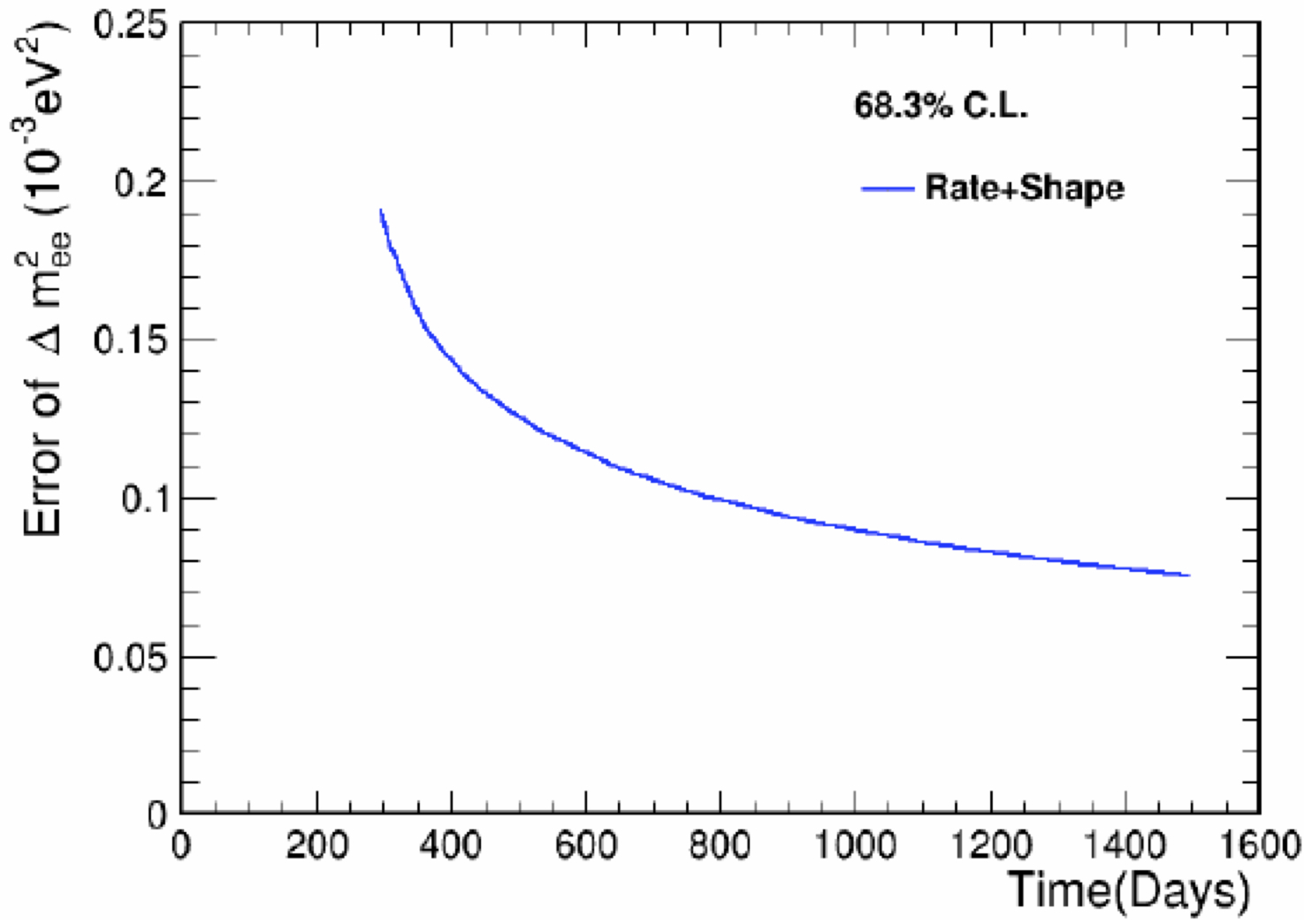}
\caption{Expected Daya Bay uncertainty in $\Delta m^2_{ee}$ as a function of running time with 8 ADs. The current systematic uncertainty is used at the beginning of the run. The present error of the 6-AD data set is $0.2\times 10^{-3}~eV^2$.}
\label{fig:ee}
\end{center}
\end{figure}

Complementary to $\Delta m^2_{\mu\mu}$ determined by accelerator-based experiments,
Daya Bay will provide a precise measurement of 
$\Delta m^2_{ee}$, as shown in Fig.~\ref{fig:ee}, better than 3\% after five years of data collection 
with the 8-AD configuration. 
In addition to the neutrino mixing angles and the Dirac CP-violating phase, the mass-squared differences are crucial for understanding the nature of neutrinos. 
First, independent determination of the three mass-squared differences will validate the sum rule of neutrino mixing, $\Delta m^2_{21}+\Delta m^2_{32}+\Delta m^2_{13}=0$. Any deviation from this sum rule could signal the existence of neutrinos beyond three generations. 
In addition, comparison of $\Delta m^2_{ee}$ and $\Delta m^2_{\mu\mu}$ will provide direct constraint to the neutrino mass hierarchy (sign of $\Delta m^2_{31}$).

\begin{figure}[htbp]
\begin{center}
\includegraphics[scale=0.6]{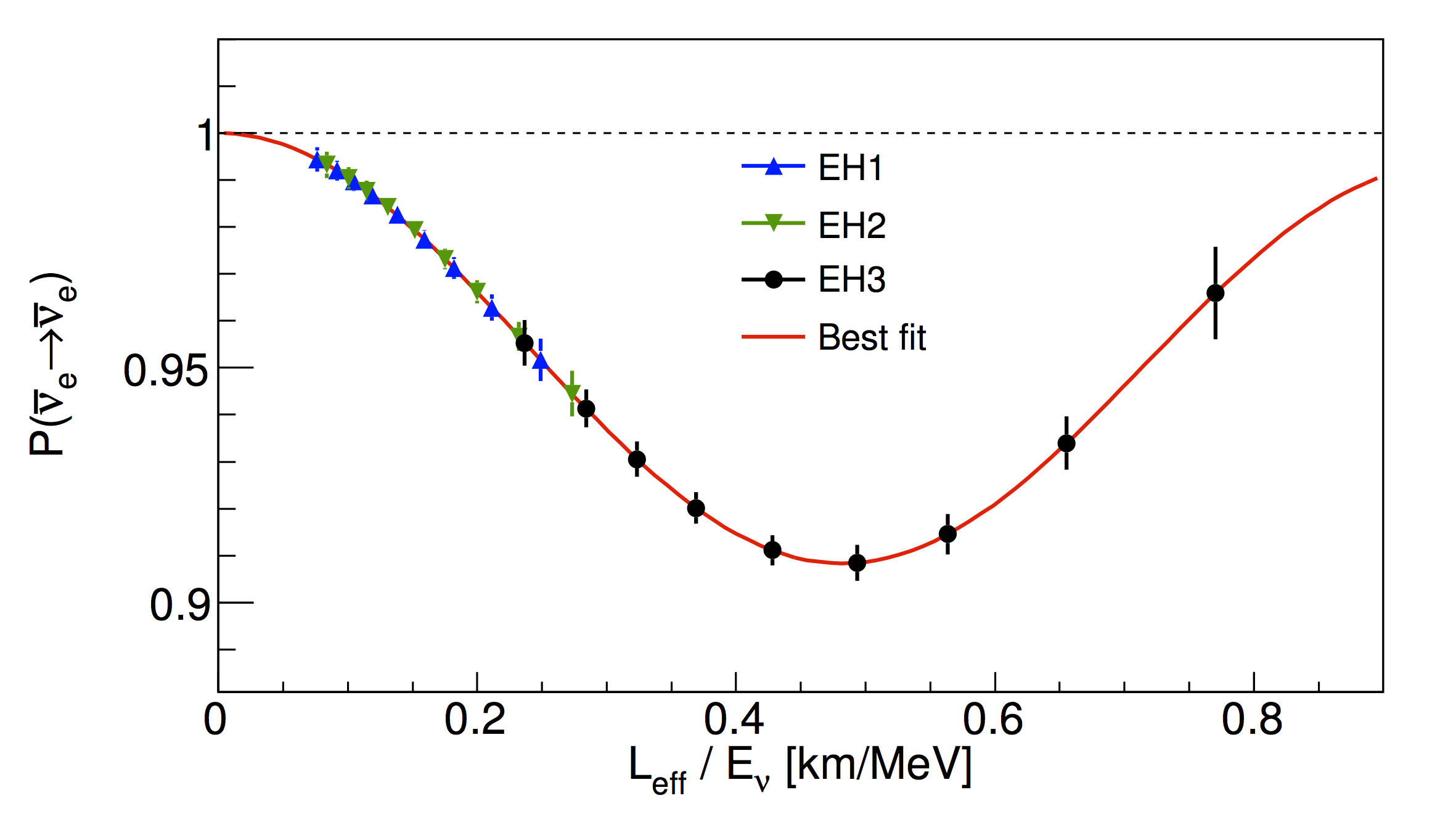}
\caption{Expected L/E distribution of Daya Bay with 8 ADs in three experimental halls, EH1,
EH2, and EH3 after three years of running. Only statistical uncertainties are shown. }
\label{fig:LoE}
\end{center}
\end{figure}

Since the fundamental parameters $\Delta m^2_{32}$ and $\Delta m^2_{31}$ are directly 
related to the effective mass splitting $\Delta m^2_{ee}$ in the L/E region for Daya Bay, the impact of the Daya Bay measurements in both sin$^2$2$\theta_{13}$ and $\Delta m^2_{ee}$ with three years of 8-AD data is illustrated in Fig.~\ref{fig:LoE}. Only statistical uncertainties are shown. The relative systematic uncertainty, which is relevant for the oscillation analysis, is still smaller than the expected statistical uncertainty of the far site. The overall normalization is not constrained in the fit, and the best-fit normalization is used in Fig.~\ref{fig:LoE}. 

\section{Precision Measurement of the Reactor Antineutrino Flux and Spectrum}\label{sec:reactor}
Daya Bay collects reactor antineutrino data at a tremendous rate which enables a precision measurement of the reactor antineutrino spectra in the four near site detectors. For the data sets published to date based on six detectors, Daya Bay has collected $>$300k antineutrino events. Two additional detectors, one at the Ling Ao near site and one at the far site, have been operational since October 2012. The Daya Bay experimental configuration allows spectral and flux measurements as close as 360m from the reactor cores at the Daya Bay site and 480m at the Ling Ao site. At these detector locations the measured reactor antineutrino flux from the nearest reactor cores remains largely non-oscillated in the standard 3-neutrino oscillation framework. The contribution from the more distant reactors at $\sim$900m is approximately 20\% (7.5\%) of the total event rate at Daya Bay (Ling Ao). The oscillation effects of antineutrinos from the far reactor and resulting distortions in the measured antineutrino spectrum can be corrected for in any spectral analysis assuming a 3-neutrino framework. With this unique configuration of multiple, functionally identical antineutrino detectors at various baselines Daya Bay will be able to report the following measurements and physics analyses:

\subsection{High-precision measurement of the reactor antineutrino spectrum}
Using data from the near and far sites Daya Bay will make a measurement of the reactor antineutrino spectrum with high precision. Amongst all running reactor experiments Daya Bay will collect the largest sample of antineutrino events  and achieve a $<$1\% statistical uncertainty in a 2-year run over a large range of energies at the near sites. 

\subsection{Test of the reactor antineutrino spectrum vs predictions and search for new interactions}
Using known reactor operation data such as thermal power output and fission fraction evolution from reactor core simulations Daya Bay can predict the expected non-oscillated energy spectrum of reactor antineutrinos emitted from each reactor. The Daya Bay collaboration works closely with the reactor company and relevant operation data are provided to the collaboration on a regular basis. A precise comparison of the predicted reactor antineutrino spectrum with the spectrum measured at Daya Bay will test our understanding and calculations of antineutrino emitted from reactors. This is particularly relevant in the context of recent discussions of our understanding of reactor flux calculations. Discrepancies in the spectral shape may point to (a) missing nuclear physics in the reactor spectrum predictions or (b) new physics beyond the 3-neutrino framework including non-standard interaction (NSI) effects, and sterile neutrinos. This spectral shape test is independent of the absolute flux normalization and the uncertainties in the predicted total antineutrino rate at Daya Bay. Due to the high statistics of the Daya Bay measurement, the statistical uncertainty in the 2011-2012 Daya Bay data set is already below the flux conversion uncertainty on the spectrum. Daya Bay's measurement of the reactor antineutrino spectrum will be ultimately limited by our understanding of energy scale uncertainties and detector effects. Detector studies and simulations are ongoing to improve the energy response model of the Daya Bay detectors.

\subsection{Absolute measurement of the reactor flux}
In addition to a measurement of the spectral shape, Daya Bay will measure the absolute reactor flux from the six Daya Bay and Ling Ao reactor cores. The absolute flux measurement tests our understanding of reactor flux models within the theoretical uncertainties of  the predictions and the experimental uncertainties associated with the absolute detection efficiencies of the Daya Bay detectors. An absolute flux measurement in the antineutrino detectors at Daya Bay will provide unique data points at the baselines of the Daya Bay experiment (360-2,000 m) and will further our understanding if there is a deficit in the measured reactor neutrino flux at short baselines, also known as the ``reactor anomaly''. An analysis of past reactor experiments compared with predictions has revealed a discrepancy of about 5.7\% in the absolute antineutrino flux. While Daya Bay has demonstrated superb relative detector uncertainties, an absolute measurement will be systematics limited. A statistical precision of 0.1\% will be achievable in the Daya Bay data set. Improvements in the analysis may eventually reduce absolute detection uncertainties to $\sim$1\%. An absolute reactor flux measurement can test  the theoretical flux prediction with the uncertainty currently estimated at about 2.7\%. We can compare the Daya Bay's measurements to previous reactor flux measurements, for example, 
the absolute Bugey-4 measurement with an uncertainty of 1.4\%. Daya Bay's measurement of the absolute flux and reactor antineutrino spectrum will provide important input to our understanding of the ``reactor anomaly''.

\subsection{Study of the time-evolution of the reactor antineutrino flux}
The large reactor antineutrino event rate measured and the unique combination of baselines and reactor cores at Daya Bay allows a detailed study of the time variation of the reactor antineutrino flux. This contains information on the operation of the reactors as well as the evolution and isotopic composition of the core's fuel. Correlating the measured antineutrino flux with reactor operations is of interest to reactor monitoring, the safeguard community, and applied neutrino science. With six reactors and 4 near-site detectors Daya Bay will provide the largest data set on reactor flux variations as a function of time.

\end{document}